\documentclass[]{spie}  

\usepackage{amsmath,amsfonts,amssymb}
\usepackage{graphicx}
\usepackage[colorlinks=true, allcolors=blue]{hyperref}

\usepackage{mathtools}
\usepackage{mathrsfs}
\makeatletter
\newcommand{\ostar}{\mathbin{\mathpalette\make@circled\star}}
\newcommand{\make@circled}[2]{%
  \ooalign{$\m@th#1\smallbigcirc{#1}$\cr\hidewidth$\m@th#1#2$\hidewidth\cr}%
}
\newcommand{\smallbigcirc}[1]{%
  \vcenter{\hbox{\scalebox{0.77778}{$\m@th#1\bigcirc$}}}%
}

\usepackage{array}
\usepackage{makecell}
\newcolumntype{C}[1]{>{\centering\arraybackslash}p{#1}} 
\newcolumntype{L}[1]{>{\raggedright\arraybackslash}m{#1}} 
\newcolumntype{M}[1]{>{\centering\arraybackslash}m{#1}} 
\newcolumntype{B}[1]{>{\centering\arraybackslash}b{#1}} 

\usepackage{graphicx}
\graphicspath{{./figures/}}

\usepackage{transparent}
\usepackage{xkeyval,xcolor}
\makeatletter
\newlength{\sfp@hseplen}\newlength{\sfp@vseplen}
\define@cmdkey{subfigpos}[sfp@]{pos}[ul]{}
\define@cmdkey{subfigpos}[sfp@]{font}[\small]{}
\define@cmdkey{subfigpos}[sfp@]{vsep}[0.75\baselineskip]{\setlength{\sfp@vseplen}{\sfp@vsep}}
\define@cmdkey{subfigpos}[sfp@]{hsep}[3.5pt]{\setlength{\sfp@hseplen}{\sfp@hsep}}
\newcommand{\subfigimg}[4][,]{%
        \setkeys{Gin,subfigpos}{pos,font,vsep,hsep,#1}
        \setbox1=\hbox{\includegraphics{#4}}
        \ifnum\pdfstrcmp{\sfp@pos}{ul}=0
                \leavevmode\rlap{\usebox1}
                \rlap{\hspace*{\sfp@hsep}\raisebox{\dimexpr\ht1-\sfp@vsep}{\transparent{#3}{\setlength{\fboxsep}{1pt}\colorbox{white}{%
\transparent{1}\sfp@font{#2}}}%
}}
                \phantom{\usebox1}
        \else\ifnum\pdfstrcmp{\sfp@pos}{ur}=0
                \leavevmode\usebox1
                \llap{\raisebox{\dimexpr\ht1-\sfp@vsep}{\sfp@font{#2}}\hspace*{\sfp@hsep}}
        \else\ifnum\pdfstrcmp{\sfp@pos}{lr}=0
                \leavevmode\usebox1
                \llap{\raisebox{\sfp@vsep}{\sfp@font{#2}}\hspace*{\sfp@hsep}}
        \else
                \leavevmode\rlap{\usebox1}
                \rlap{\hspace*{\sfp@hseplen}\raisebox{\sfp@vsep}{\sfp@font{#2}}}
                \phantom{\usebox1}
        \fi\fi\fi
}
\newcommand{\fontfig}[1]{\small$\!\!$\color{#1}\textbf}
\newcommand{\AspectRatio}[1]{\dimexpr 1pt * \wd#1 / \ht#1 \relax} 

\renewcommand{\refeq}[1]{Eq.~(\ref{#1})\xspace} 
\newcommand{\refeqs}[2]{Eqs.~(\ref{#1}) and (\ref{#2})\xspace}
\newcommand{\refeqfull}[1]{Equation~(\ref{#1})\xspace} 

\newcommand{\reffig}[1]{Fig.~\ref{#1}\xspace}
\newcommand{\reffigs}[2]{Figs.~\ref{#1} and~\ref{#2}\xspace} 
\newcommand{\reffigfull}[1]{Figure~\ref{#1}\xspace} 

\newcommand{\refsub}[1]{#1} 
\newcommand{\refsubfig}[2]{Fig.~\ref{#1}\refsub{#2}\xspace}
\newcommand{\refsubfigfull}[2]{Figure~\ref{#1}\refsub{#2}\xspace} 
\newcommand{\refsubfigs}[2]{Figs.~\ref{#1}\refsub{#2}\xspace}

\newcommand{\refpan}[1]{Panel~#1} 
\newcommand{\refpans}[1]{Panels~#1} 

\newcommand{\reftab}[1]{Table~\ref{#1}\xspace}

\newcommand{\refsec}[1]{Sec.~\ref{#1}\xspace} 
\newcommand{\refsecs}[2]{Secs.~\ref{#1} and~\ref{#2}\xspace} 

\newcommand{\citet}[1]{Ref.~\citenum{#1}\xspace}

\DeclarePairedDelimiterX{\paren}[1]{(}{)}{#1}
\newcommand{\Paren}[1]{\paren*{#1}}
\let\brace=\undefined 
\DeclarePairedDelimiterX{\brace}[1]{\{}{\}}{#1}

\let\brack=\undefined 
\DeclarePairedDelimiterX{\brack}[1]{[}{]}{#1}
\newcommand{\Brack}[1]{\brack*{#1}}
\DeclarePairedDelimiterX{\bbrack}[1]{\llbracket}{\rrbracket}{#1}

\DeclarePairedDelimiterX{\abs}[1]{\rvert}{\lvert}{#1}     
\newcommand{\Abs}[1]{\abs*{#1}}
\DeclarePairedDelimiterX{\norm}[1]{\lVert}{\rVert}{#1}    

\DeclarePairedDelimiterX{\avg}[1]{\langle}{\rangle}{#1}   
\newcommand{\Avg}[1]{\avg*{#1}}

\DeclarePairedDelimiterX{\ceil}[1]{\lceil}{\rceil}{#1}     

\DeclarePairedDelimiterX{\floor}[1]{\lfloor}{\rfloor}{#1}  

\newcommand{\crosscorr}{\ostar}                         

\newcommand{\imag}[1]{\mathscr{I}\Brack{#1}} 

\newcommand{\Tag}[1]{\text{#1}}                 

\newcommand{\V}[1]{{\boldsymbol{#1}}}   
\newcommand{\conj}[1]{\overline{#1}}    
\newcommand{\scaprod}{^T\!\!\cdot}            

\newcommand{\x}{{x}}
\newcommand{\Vx}{{\V{\x}}}
\newcommand{\Vk}{{\V{k}}} 
\newcommand{\Vv}{{\V{v}}} 
\newcommand{\Vdelta}{{\V{\delta}}} 

\newcommand{\Vm}{\V{m}}
\newcommand{\Vw}{\V{w}}
\newcommand{\Vn}{\V{n}}
\newcommand{\Vc}{\V{c}}
\newcommand{\Vp}{\V{p}}

\newcommand{\muconj}{\conj{\mu}}

\newcommand{\Vvar}[1]{\V{\mathscr{V}}_{\!\!\!#1}} 
\newcommand{\corr}[1]{\mathscr{C}_{\!#1}} 
\newcommand{\Vcorr}[1]{\V{\mathscr{C}}_{\!#1}} 
\newcommand{\corrt}[1]{\tilde{\mathscr{C}}^\Tag{#1}} 
\newcommand{\Vcorrt}[1]{\tilde{\V{\mathscr{C}}}^\Tag{#1}} 
\newcommand{\argmin}[2]{\underset{#1}{\text{argmin}}\;#2}
\newcommand{\argmax}[2]{\underset{#1}{\text{argmax}}\;#2}

\newcommand{\CM}{\V{M}^\Tag{cmd}}
\newcommand{\IM}[1]{\V{M}^\Tag{int}_{#1}}
\newcommand{\IMtilde}[1]{\tilde{\V{M}}^\Tag{int}_{#1}}

\newcommand{\rFried}{r_{0}}
\newcommand{\vWind}{v_{0}}

\newcommand{\SApitch}{\Delta}
\newcommand{\nmodes}{n^\Tag{mod}}

\newcommand{\kctrl}{\mathcal{K}^\Tag{ctrl}}
\newcommand{\fwind}{f^\parallel}

\newcommand{\weight}{{w}}
\newcommand{\weightWFS}{\weight^{\Tag{wfs}}}
\newcommand{\weightvalid}{\weight^{\Tag{valid}}}

\newcommand{\tDM}{\tau^{\Tag{dm}}}
\newcommand{\tlat}{\tau^{\Tag{lat}}}
\newcommand{\tRTC}{\tau^{\Tag{rtc}}}
\newcommand{\tWFS}{\tau^{\Tag{wfs}}}
\newcommand{\gint}{g^{\int}}
\newcommand{\gleak}{g^{\epsilon}}

\newcommand{\X}[1]{$#1\!\times\!#1$}

\usepackage{siunitx}
\newcommand{\percent}[1]{#1\,\%}     

\usepackage{xspace}

\title{Open loop calibration and closed loop non perturbative estimation of the lateral errors of an adaptive optics system: examples with GRAVITY+ and CHARA experimental data}

\author[a]{Anthony Berdeu}
\author[b]{Henri Bonnet}
\author[c]{Jean-Baptiste Le Bouquin}
\author[b]{Johann Kolb}
\author[d]{Guillaume Bourdarot}
\author[e]{Philippe Berio}
\author[a]{Thibaut Paumard}
\author[d]{Frank Eisenhauer}
\author[f]{Christian Straubmeier}
\author[g,h]{Paulo Garcia}
\author[i]{Sebastian Hönig}
\author[d]{Florentin Millour}
\author[j]{Laura Kreidberg}
\author[k]{Denis Defrère}
\author[l]{Ferréol Soulez}
\author[d]{Denis Mourard}
\author[m]{Gail Schaefer}
\author[m]{Narsireddy Anugu}

\affil[a]{LESIA, Observatoire de Paris, Université PSL, Sorbonne Université, Université Paris Cité, CNRS, 5 place Jules Janssen, 92195 Meudon, France}
\affil[b]{European Southern Observatory, Karl-Schwarzschild-Straße 2, 85748 Garching, Germany}
\affil[c]{Univ. Grenoble Alpes, CNRS, IPAG, 38000 Grenoble, France}
\affil[d]{Max Planck Institute for extraterrestrial Physics, 85748 Garching, Germany}
\affil[e]{Université Côte d’Azur, Observatoire de la Côte d’Azur, CNRS, Laboratoire Lagrange, France}
\affil[f]{1st Institute of Physics, University of Cologne, Zülpicher Straße 77, 50937 Cologne, Germany}
\affil[g]{Centro de Astrofísica e Gravitação, IST, Universidade de Lisboa, 1049-001 Lisboa, Portugal}
\affil[h]{Faculdade de Engenharia, Universidade do Porto, 4200-465 Porto, Portugal}
\affil[i]{School of Physics \& Astronomy, University of Southampton, Southampton, SO17 1BJ, UK}
\affil[j]{Max Planck Institute for Astronomy, Königstuhl 17, 69117 Heidelberg, Germany}
\affil[k]{Institute of Astronomy, KU Leuven, Celestijnenlaan 200D, B-3001, Leuven, Belgium}
\affil[l]{Univ Lyon, Univ Lyon1, Ens de Lyon, Centre de Recherche Astrophysique de Lyon, UMR 5574, F-69230, Saint-Genis-Laval, France}
\affil[m]{The CHARA Array of Georgia State University, Mount Wilson Observatory, Mount Wilson, Altadena, CA 91203, United States of America}

\authorinfo{On the behalf of the GRAVITY+ Collaboration. Send correspondence to Anthony Berdeu: \href{mailto:anthony.berdeu@obspm.fr}{anthony.berdeu@obspm.fr}}

\begin{document} 
\maketitle

\begin{abstract}
Performances of an adaptive optics (AO) system are directly linked with the quality of its alignment. During the instrument calibration, having open loop fast tools with a large capture range are necessary to quickly assess the system misalignment and to drive it towards a state allowing to close the AO loop. During operation, complex systems are prone to misalignments (mechanical flexions, rotation of optical elements, ...) that potentially degrade the AO performances, creating a need for a monitoring tool to tackle their driftage. In this work, we first present an improved perturbative method to quickly assess large lateral errors in open loop. It uses the spatial correlation of the measured interaction matrix of a limited number of 2D spatial modes with a synthetic model. Then, we introduce a novel solution to finely measure and correct these lateral errors via the closed loop telemetry. Non-perturbative, this method consequently does not impact the science output of the instrument. It is based on the temporal correlation of 2D spatial frequencies in the deformable mirror commands. It is model-free (no need of an interaction matrix model) and sparse in the Fourier space, making it fast and easily scalable to complex systems such as future extremely large telescopes. Finally, we present some results obtained on the development bench of the GRAVITY+ extreme AO system (Cartesian grid, 1432 actuators). In addition, we show with on-sky results gathered with CHARA and GRAVITY/CIAO that the method is adaptable to non-conventional AO geometries (hexagonal grids, 60 actuators).
\end{abstract}


\section{INTRODUCTION}
\label{sec:intro}

The role of an Adaptive Optics (AO) system is to compensate for the wavefront corrugations induced by atmospheric turbulence\cite{Roddier:99_AO_system}. To do so, such a system is composed by a wavefront sensor (WFS) whose measurements are analyzed by a real time computer (RTC) and converted into a set of commands sent to a deformable mirror (DM) that corrects the optical aberrations. To be effective, this feedback loop must run faster than the turbulence temporal evolution, with typical frequencies ranging from several hundred Hertz to a kilo-Hertz.

Any alignment error in the AO system degrades its performances, or even worse, destabilizes the AO loop. Fine registration of the couple DM/WFS is thus critical. Among others, see \citet{Heritier:19_PHD} (Sec. 2.1), the classical (and most impacting) mis-registrations are the $x$ and $y$-shifts (lateral translation of the DM with respect to the WFS), the clocking (rotation of the DM with respect to the WFS), and the magnification and anamorphosis (stretches of the DM with respect to the WFS).

If different techniques exist to assess the mis-registration state of an AO system\cite{Kolb:16_SPIE_Review_AO_calibration}, most of the tools to track mis-registrations use the interaction matrix (IM) of the AO system\cite{Roddier:99_AO_system} that links the commands sent to the DM to the measures provided by the WFS. These techniques rely on the fact that an alignment error in the AO system leaves a measurable signal in its IM.

In early AO systems, IMs were experimentally calibrated on instrument internal sources with high signal-to-noise ratios (S/Ns)\cite{Oberti:06_SPIE_PSIM_vs_measured_IM}. But the arrival of telescopes with adaptive secondary mirrors, such as the Multiple Mirror Telescope\cite{Brusa:03_MMT}, the Large Binocular Telescope\cite{Riccardi:03_SPIE_LBT_SDM}, or the AOF\cite{Strobele:06_SPIE_AOF}, changed the deal. Without any reference source, the IM must be measured directly on-sky at night. This raised new challenges\cite{Oberti:06_SPIE_PSIM_vs_measured_IM, Pinna:12_SPIE_FLAO, Lai:21_DO_CRIME,Kolb:16_SPIE_Review_AO_calibration}: freezing the turbulence, fighting low S/N, minimizing the loss of precious night time for the scientific observations, ...

Thus, new methods emerged, based on the physical modeling of the interaction matrix, so-called synthetic IM\cite{Oberti:06_SPIE_PSIM_vs_measured_IM, Kolb:12_SPIE_AOF, Heritier:18_PyWFS_calibration, Heritier:19_PHD}. The mis-registration parameters are fitted through dedicated calibrations, on internal sources or on-sky\cite{Oberti:04_SPIE_IM_misreg, Neichel:12_SPIE_MCAO, Kolb:16_SPIE_Review_AO_calibration, Heritier:18_PyWFS_calibration} to obtain noiseless pseudo-synthetic interaction matrices\cite{Oberti:06_SPIE_PSIM_vs_measured_IM} (PSIMs).

Nonetheless, mis-registrations are susceptible to evolve during observation (mechanical flexion, thermal evolution, ...). And future Extremely Large Telescopes\cite{Johns:06_SPIE_GMT, Gilmozzi:07_ELT, Boyer:18_SPIE_TMT}(ELTs) will bring this challenge to other levels: always more complex (and time consuming) PSIM models, increasing number of actuators, non-linear WFSs, unprecedented distances between DMs and WFSs (with moving/rotating parts in-between prone to misalignment), ... It then becomes impossible to perform regular calibrations and it is thus necessary to track the evolution of the mis-registrations directly during the scientific acquisitions with online tools for AO system auto-calibration.

In this context we developed two news methods to assess lateral mis-registration in an AO system that relax the need of complex PSIM model fitting. The first one, see \refsec{sec:method_2D_corr}, is based on a perturbative approach whereby two-dimensional (2D) modes are applied on the system. It has a large capture range and can be used in open loop in order to calibrate and preset the AO system. The second one, see \refsec{sec:method_CL_estim}, is a non-perturbative approach analyzing correlation signals in the AO loop telemetry. Working in closed loop, it can be used in real time to monitor the system mis-registration without impacting the science acquisition.

These methods have been introduced in a previous paper, see \citet{Berdeu:24_misreg} and implemented in the Standard Platform for Adaptive optics Real Time Applications\cite{Suarez:12_SPIE_SPARTA, Shchekaturov:23_SPARTA_upgrade, Dembet:23_RTC_GPAO} (SPARTA), the RTC of GRAVITY+\cite{GRAVITYplus:22_messenger}. In this work, we present new data, obtained on the GRAVITY+ AO development bench\cite{Millour:22_SPIE_GPAO_bench, LeBouquin:23_GPAO_design}(GPAO), as well as the first on-sky results obtained from the Center for High Angular Resolution Astronomy\cite{Anugu:2020_CHARA_SPIE} (CHARA) and the Coudé Infrared Adaptive Optics\cite{Kendrew:12_CIAO} (CIAO) of GRAVITY\cite{GRAVITY:17_VLTI}. For each technique, (i)~a brief reminder of the method is introduced, see \refsecs{sec:meth_2D_corr}{sec:meth_CL_estim}; (ii)~then experimental results are given, see \refsecs{sec:exp_2D_corr}{sec:exp_CL_estim}.

\section{Perturbative 2D modal estimator}
\label{sec:method_2D_corr}

\subsection{Overview of the Method}
\label{sec:meth_2D_corr}

\subsubsection{General concept}

\begin{figure}[t!]
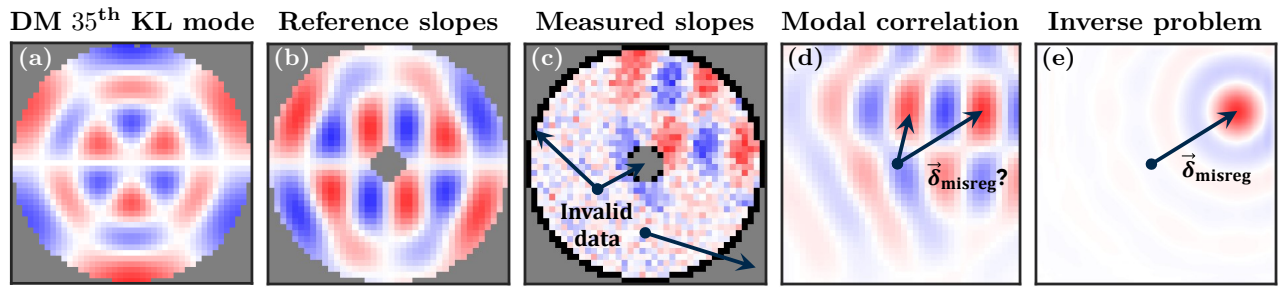
 
        \centering
        
        \newcommand{\PathFig}{Fig_misreg_2D_sim/}
        
        \newcommand{\LineRatio}{0.95}
        
        \newcommand{\fontTxt}[1]{\textbf{#1}}
        
        \newcommand{\subfigColor}{black}
        
        \newcommand{\figOne}{\PathFig Mode_DM}
        \newcommand{\figTwo}{\PathFig Mode_reference}
        \newcommand{\figThree}{\PathFig Mode_measured}
        \newcommand{\figFour}{\PathFig Mode_correlation}
        \newcommand{\figFive}{\PathFig Mode_inverse_problem}
        
        \sbox1{\includegraphics{\figOne}}
        \sbox2{\includegraphics{\figTwo}}
        \sbox3{\includegraphics{\figThree}}
        \sbox4{\includegraphics{\figFour}}
        \sbox5{\includegraphics{\figFive}}
        
        \newcommand{\ColumnWidth}[1]
                {\dimexpr \LineRatio \linewidth * \AspectRatio{#1} / (\AspectRatio{1} + \AspectRatio{2} + \AspectRatio{3} + \AspectRatio{4} + \AspectRatio{5}) \relax
                }
        \newcommand{\ColumnGap}{\hspace {\dimexpr \linewidth /6 - \LineRatio\linewidth /6 }}

        \begin{tabular}{
                @{\ColumnGap}
                M{\ColumnWidth{1}}
                @{\ColumnGap}
                M{\ColumnWidth{2}}
                @{\ColumnGap}
                M{\ColumnWidth{3}}
                @{\ColumnGap}
                M{\ColumnWidth{4}}
                @{\ColumnGap}
                M{\ColumnWidth{5}}
                @{\ColumnGap}
                }
                
                \fontTxt{DM $35^\Tag{th}$ KL mode}
                &
                \fontTxt{Reference slopes}
                &
                \fontTxt{Measured slopes}
                &
                \fontTxt{Modal correlation}
                &
                \fontTxt{Inverse problem}
                \\                
                
                \subfigimg[width=\linewidth,pos=ul,font=\fontfig{white}]{$\,$(a)}{0.0}{\figOne} &
                \subfigimg[width=\linewidth,pos=ul,font=\fontfig{white}]{$\,$(b)}{0.0}{\figTwo} &
                \subfigimg[width=\linewidth,pos=ul,font=\fontfig{white}]{$\,$(c)}{0.0}{\figThree} &
                \subfigimg[width=\linewidth,pos=ul,font=\fontfig{\subfigColor}]{$\,$(d)}{0.0}{\figFour}&
                \subfigimg[width=\linewidth,pos=ul,font=\fontfig{\subfigColor}]{$\,$(e)}{0.0}{\figFive}
                \\
        \end{tabular}        
               
        \caption{\label{fig:2D_corr_sim_misreg} Simulated example of the spatial 2D correlation method. \refpan{a}: KL mode applied on the DM. \refpan{b}: Corresponding reference slopes on the WFS. \refpan{c}: Slopes of the measured partial IM. \refpan{d}: Standard correlation of the measured KL mode with the reference. \refpan{e}: Global inverse problem approach on all the measured modes.
        }
\end{figure}

The idea of the method is summed up in \reffig{fig:2D_corr_sim_misreg}. When a spatial pattern is applied on the DM, see \refsubfig{fig:2D_corr_sim_misreg}{a}, a specific spatial pattern is seen in the SH-WFS data, see \refsubfig{fig:2D_corr_sim_misreg}{b}. In the following, we used a modal approach based on Karhunen–Loève (KL) functions\cite{Dai:95_KL} applied on the DM commands (modal IM). If the considered modes are properly sampled, a lateral mis-registration geometrically shifts the measurements, see \refsubfig{fig:2D_corr_sim_misreg}{c}. Thus, the 2D spatial correlation of a measured modal IM with a reference modal IM should be usable as a lateral error estimator.

Noting $\IM{m}\Paren{\Vx}$ (resp. $\IMtilde{m}\Paren{\Vx}$) the reference modal IM of the system without any mis-registration (resp. the measured modal IM) of the KL mode $m$, the cross-correlation of the 2D IMs for a shift~$\Vdelta$ is given by:
\begin{equation}
    \label{eq:cross_corr}
    \Brack{\IMtilde{m}\crosscorr\IM{m}}\Paren{\Vdelta}
    {} \triangleq {} 
    \sum_{\Vx}\IMtilde{m}\Paren{\Vx}\IM{m}\Paren{\Vx-\Vdelta}
    {} = {}  \alpha\Paren{\Vdelta}
    \,.
\end{equation}
In presence of a lateral mis-registration, \refeq{eq:cross_corr} would give maps similar to \refsubfig{fig:2D_corr_sim_misreg}{d}. Performed on a single mode, such correlation maps can lead to ambiguities in the determination of the maximum location and to poor sensitivity. In our method, we rather define the merit criterion in terms of how similar the two IMs are in terms of mean squares, jointly accounting for all the modes,
\begin{equation}
    \label{eq:2D_mod_min}
    \alpha\Paren{\Vdelta} = \argmin{\beta}{\sum_{\Vx,m}w\Paren{\Vx,\Vdelta}\Paren{\IMtilde{m}\Paren{\Vx} - \beta\IM{m}\Paren{\Vx-\Vdelta}}^2}
    \,,
    \text{ with }
    w\Paren{\Vx,\Vdelta} = \weightvalid\Paren{\Vx}\weightWFS\Paren{\Vx-\Vdelta}
    \,,
\end{equation}
where $\weightWFS\Paren{\Vx}$ is the map of the valid subapertures in the PSIM model\footnote{Such invalid subapertures are in gray in \refsubfig{fig:2D_corr_sim_misreg}{c}.} and $\weightvalid\Paren{\Vx}$ is the map of the valid subapertures in the measurements, discarding poorly illuminated subapertures\footnote{Such invalid subapertures are in black in \refsubfig{fig:2D_corr_sim_misreg}{c}.}. \refeqfull{eq:2D_mod_min} has an analytical solution given by:
\begin{equation}
    \label{eq:2D_corr_map}
    \alpha\Paren{\Vdelta} {}={} \frac{\sum_{\Vx,m}w\Paren{\Vx,\Vdelta}\IMtilde{m}\Paren{\Vx}\IM{m}\Paren{\Vx-\Vdelta}}{\sum_{\Vx,m}w\Paren{\Vx,\Vdelta}\Brack{\IM{m}\Paren{\Vx-\Vdelta}}^2}
    {}={} \frac{\sum_{m}\Brack{\weightvalid\IMtilde{m}\crosscorr\weightWFS\IM{m}}\Paren{\Vdelta}}{\sum_{m}\Brack{\weightvalid\crosscorr\weightWFS\Brack{\IM{m}}^2}\Paren{\Vdelta}}
    \,.
\end{equation}
\refeqfull{eq:2D_corr_map} is the ratio of the modal sum of 2D cross-correlations and the position of the global maximum of the obtained map, see \refsubfig{fig:2D_corr_sim_misreg}{e}, gives the estimated lateral mis-registration~$\tilde{\Vdelta}$ in a single pass:
\begin{equation}
    \label{eq:2D_corr_estim}
    \tilde{\Vdelta} = \argmax{\Vdelta}\alpha\Paren{\Vdelta}
    \,.
\end{equation}
As discussed in \citet{Berdeu:24_misreg}, this map can be up-sampled to perform super-resolution.

\subsection{Experimental Results}
\label{sec:exp_2D_corr}

\begin{figure}[t!]
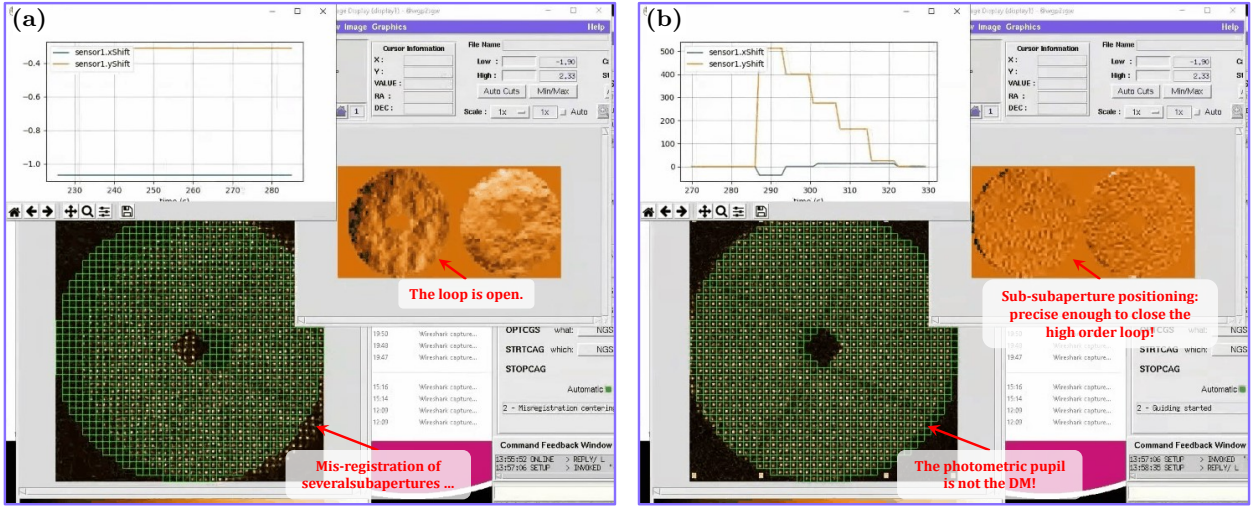
 
        \centering
        
        \newcommand{\PathFig}{Fig_misreg_2D_bench/}
        
        \newcommand{\LineRatio}{0.95}
        
        \newcommand{\subfigColor}{black}
        
        \newcommand{\figOne}{\PathFig 2D_corr_init}
        \newcommand{\figTwo}{\PathFig 2D_corr_conv}
        
        \sbox1{\includegraphics{\figOne}}
        \sbox2{\includegraphics{\figTwo}}
        
        \newcommand{\ColumnWidth}[1]
                {\dimexpr \LineRatio \linewidth * \AspectRatio{#1} / (\AspectRatio{1} + \AspectRatio{2}) \relax
                }
        \newcommand{\ColumnGap}{\hspace {\dimexpr \linewidth /3 - \LineRatio\linewidth /3 }}

        \begin{tabular}{
                @{\ColumnGap}
                M{\ColumnWidth{1}}
                @{\ColumnGap}
                M{\ColumnWidth{2}}
                @{\ColumnGap}
                }
                
                \subfigimg[width=\linewidth,pos=ul,font=\fontfig{\subfigColor}]{$\,$(a)}{1.0}{\figOne} &
                \subfigimg[width=\linewidth,pos=ul,font=\fontfig{\subfigColor}]{$\,$(b)}{1.0}{\figTwo}
                \\
        \end{tabular}        
               
        \caption{\label{fig:2D_corr_bench} Screenshots of the spatial 2D correlation method in the GPAO bench.  The orange and blue curves are the $x$ and $y$ outputs of the estimator according to time. The SH-WFS spots are displayed in the bottom left and the corresponding $x$ and $y$ slopes are given side by side in the windows with the orange look up table. \refpan{a}: Initial state of the AO system with a lateral error of several subapertures. \refpan{b}: System after convergence of the lateral error corrective loop.
        }
\end{figure}

\reffigfull{fig:2D_corr_bench} gathers some screenshots of the SPARTA panels that show the use of the proposed method during the present of the visible mode of GPAO, based on a \X{40} Shack-Hartmann WFS (SH-WFS) and a DM with 1432 active actuators. The turbulence was emulated with a plate, producing a wind of $\vWind\simeq\SI{8.4}{\meter\per\second}$ with a Fried parameter of~$\rFried \simeq \SI{14}{\centi\meter}$.

In the initial state, see \refsubfig{fig:2D_corr_bench}{a}, the loop is open and the system is misaligned by several subapertures. This is visible by the important shift between the photometric pupil (SH-WFS spots) relative to the SH-WFS geometry (green boxes). \refsubfigfull{fig:2D_corr_bench}{b} shows the system after the convergence of the corrective loop that monitors the lateral error based on the 2D modal estimator. The system has converged in a few iterations and the alignment is good enough to close the high order AO loop, with 800 controlled modes.

As emphasized by the red arrow, an alignment of the DM with respect to the WFS does not correspond to the alignment of the photometric pupil. This is explained by the fact that in GPAO, the DM clear aperture is larger than the telescope pupil image and thus photometry centering cannot be used to register the couple DM/WFS.

The algorithm has been tested in the bench with the different GPAO modes (\X{9}, \X{30}, \X{40} SH-WFSs) and with different phase plates and source powers, spanning a large variety of conditions. In all cases, it successfully brings the system to a state where the high order AO loop closes stably. The mis-registration monitoring can thus then switch to the closed loop estimator, as discussed below.

\section{Non-perturbative closed loop estimator}
\label{sec:method_CL_estim}

\subsection{Overview of the Method}
\label{sec:meth_CL_estim}

As summed by the blue (or green) block diagram of \refsubfig{fig:CL_corr_curves}{a}, an AO loop works as follows:
\begin{itemize}
	\item[(1)] the WFS~$S$ converts the 2D wavefront~$\Vw$ into measurements~$\Vm$ corrupted by noise,~$\Vn$: $\Vm = S\Paren{\Vw} + \Vn$,
	\item[(2)] these measurements are converted via $\CM$, the command matrix (computed by inverting the PSIM of the system filtered on a given~$\nmodes$ number of KL modes), to a command correction $\Vc^{\delta} = \CM \Vm$,
	\item[(3)] the controller~$C$, for example a leaky integrator of leak gain~$\gleak$ and integral gain~$\gint$, computes a new command $\Vc^{i+1} = \Paren{1-\gleak}\Vc^{i} + \gint\Vc^{\delta}$,
	\item[(4)] this command is applied on the DM actuators~$C$ until the next command arrives.
\end{itemize}
All the above steps are linear. Thus, the AO loop can be described in the Fourier space of the commands~$\Vm$, that is to say in terms of spatial frequencies~$\Vk=\Paren{k_x, k_y}$~(\SI{}{\per\meter}) propagating through the system. The transfer functions\cite{Astrom:21_Feedback_system} in the temporal frequency space~$f$ (\SI{}{\hertz}) of each block of \refsubfig{fig:CL_corr_curves}{a} are given by\cite{Madec:99_control}
\begin{itemize}
	\item $S\Paren{f} = \frac{1-e^{-2i\pi\tWFS f}}{2i\pi\tWFS f}$, where $\tWFS$ is the WFS exposure time,
	\item $C\Paren{f} = \frac{\gint e^{-2i\pi\tlat f}}{1-\Paren{1-\gleak} e^{-2i\pi\tRTC f}}$, where~$\tlat$ is the latency of the system (communication and computation times) and~$\tRTC$ is the period of the RTC cycle,
	\item $A\Paren{f} = \frac{1-e^{-2i\pi\tDM f}}{2i\pi\tDM f}$, where $\tDM$ is the holding time of the DM.
\end{itemize}
In general, all the characteristic times are equal: $\tlat \simeq \tDM \simeq \tWFS \simeq \tRTC$.

\begin{figure}[t!] 
        \centering
        
        \newcommand{\PathFig}{Fig_misreg_CL_corr_curves/}
        
        \newcommand{\LineRatio}{0.85}
        
        \newcommand{\subfigColor}{black}
        
        \newcommand{\figOne}{\PathFig Fig_spatial_coupling_SPIE}
        \newcommand{\figTwo}{\PathFig corr_noise}
        \newcommand{\figThree}{\PathFig corr_wind}
        
        \sbox1{\includegraphics{\figOne}}
        \sbox2{\includegraphics{\figTwo}}
        \sbox3{\includegraphics{\figThree}}
        
        
        \newcommand{\ColumnWidth}[1]
                {\dimexpr \LineRatio \linewidth * \AspectRatio{#1} / (\AspectRatio{1} + \AspectRatio{2}) \relax
                }
        \newcommand{\ColumnGap}{\hspace {\dimexpr \linewidth /3 - \LineRatio\linewidth /3 }}

        \begin{tabular}{
                @{\ColumnGap}
                M{\ColumnWidth{1}}
                @{\ColumnGap}
                M{\ColumnWidth{2}}
                @{\ColumnGap}
                }
                
                \subfigimg[width=\linewidth,pos=ul,font=\fontfig{\subfigColor}]{$\!\!\!\!\!\!\!\!$(a)}{0.5}{\figOne} &
                \subfigimg[width=\linewidth,pos=ul,font=\fontfig{\subfigColor}]{$\!\!\!\!$(b)}{0.5}{\figTwo}
                \\
        \end{tabular}      
               
        \caption{\label{fig:CL_corr_curves} Temporal coupling of the spatial frequency $\Vk=\Paren{2/3D, 0}$ in the AO loop telemetry. \refpan{a}: Block diagram of the temporal coupling (in red) of the symmetric (cosine, in blue) and anti-symmetric (sine, in green) parts. \refpan{b}: Imaginary part of the correlation curve~$\Vcorr{\Vc_{1},\Vc_{2}}\Paren{\theta,f}$ in a noise limited regime with standard AO parameters~$\tWFS=\SI{1}{\milli\second}$, $\gint=0.5$, $\gleak=0$ and for various values of coupling coefficient $\theta$ from \SI{0}{\degree} (blue) to \SI{45}{\degree} (red), every \SI{5}{\degree}.
        } 
\end{figure}

In the presence of a lateral mis-registration $\Vdelta=\Paren{\delta_{x}, \delta_{y}}$, the spatially symmetric part of the commands~$\Vc_{1}$ is partially projected on the spatially anti-symmetric part of the WFS measurements and propagates to the anti-symmetric part of the command~$\Vc_{2}$ (and oppositely). This cross-coupling is emphasized by the red connectors in \refsubfig{fig:CL_corr_curves}{a}, with a coupling coefficient of:
\begin{equation}
    \label{eq:theta}
    \theta = 2\pi\Vk\scaprod\Vdelta
    \,.
\end{equation}
Thus, the commands~$\Vc_{1}$ and~$\Vc_{2}$ are thus no longer independent. In the (spatial and temporal) Fourier space, the correlation a given spatial frequency~$\Vk$ at a temporal frequency~$f$ is defined by:
\begin{equation}
    \label{eq:corr}
    \Vcorr{\Vc_{1},\Vc_{2}}\Paren{\Vk,f} \triangleq \frac{\Avg{c_{1}\Paren{\Vk,f}\conj{c}_{2}\Paren{\Vk,f}}}{\sqrt{\Vvar{c_{1}}\Paren{\Vk,f}\Vvar{c_{2}}\Paren{\Vk,f}}}
    \text{ with }
    \Vvar{\Vc_{i}}\Paren{\Vk,f} \triangleq \Avg{c_{i}\Paren{\Vk,f}\conj{c}_{i}\Paren{\Vk,f}}
    \,.
\end{equation}
Under the assumption that the different noise terms~$\Vn_{i}$ and~$\Vp_{i}$ are independent, we showed in \citet{Berdeu:24_misreg} that:
\begin{equation}
    \label{eq:corr_0}
    \Vcorr{\Vc_{1},\Vc_{2}}\Paren{\theta\Paren{\Vk},f}  / \theta\Paren{\Vk} \underset{\theta\rightarrow0}{\sim} 2i\imag{\frac{\muconj\Paren{f}}{1+\muconj\Paren{f}}}
    \triangleq i\Vcorr{0}\Paren{f}
    \text{ with }
    \mu \triangleq A C S
    \,.
\end{equation}
The correlation of a given spatial frequency~$\Vk$ is thus a pure imaginary number. \refsubfigfull{fig:CL_corr_curves}{b} shows the imaginary parts of the correlation~$\Vcorr{\Vc_{1},\Vc_{2}}\Paren{\theta,f}$ for different values of the coupling coefficient~$\theta$.

On the opposite side, in a pure frozen flow hypothesis (no noise propagation), of velocity $\Vv$, it comes with $\fwind = \Vk\scaprod\Vv$:
\begin{equation}
    \forall \Abs{f} \neq \fwind, \Avg{p_{1}\conj{p}_2}\Paren{f} = 0
    \text{ / }
    f=\fwind, \corr{c_1,c_2}\Paren{\fwind} \underset{\theta\rightarrow0}{\sim} i
    \text{ / }
    f=-\fwind, \corr{c_1,c_2}\Paren{-\fwind} \underset{\theta\rightarrow0}{\sim} -i
    \,.
\end{equation}
This means that a frozen flow turbulence has a minimal impact on the correlation and only frequencies close to $f\simeq\Vk\scaprod\Vv$ should be impacted.

In practice, as detailed in \citet{Berdeu:24_misreg}, the symmetric ($\Vc_{1}$) and anti-symmetric ($\Vc_{2}$) parts of the commands~$\Vc$ are extracted from the closed loop telemetry. Then, the imaginary part of the empirical correlation of the closed loop telemetry is estimated by:
\begin{equation}
    \label{eq:corr_empirical}
    \corrt{cl}\Paren{\Vk,f} = \imag{\frac{c_{1}\Paren{\Vk,f}\conj{c}_{2}\Paren{\Vk,f}}{\Abs{c_{1}\Paren{\Vk,f}}\Abs{c_{2}\Paren{\Vk,f}}}}
    \,.
\end{equation}
And finally, the lateral error is then obtained from \refeqs{eq:theta}{eq:corr_0} by solving:
\begin{equation}
    \label{eq:CL_estim}
    \tilde{\Vdelta} = \argmin{\Vdelta}{\sum_{\Vk\in\kctrl, f>0}\Paren{\corrt{cl}\Paren{\Vk,f} - 2\pi\Vcorr{0}\Paren{f}\Vk\scaprod\Vdelta}^2}
    \,,
\end{equation}
where~$\Vk\in\kctrl$ is the control space of the AO loop.

\subsection{Experimental Results}

\label{sec:exp_CL_estim}

\subsubsection{On the GPAO bench}

The estimator was tested in the GPAO bench inserting two phase screens in the bench to emulate a multi-layered atmosphere with an estimated global seeing of $\sim 1.1''$. In the initial state, a lateral mis-registration of \percent{60} of a subaperture $\SApitch$ was injected in the system by translating the SH-WFS stage.

\refsubfigfull{fig:CL_corr_GPAO}{a} shows the convergence curve of the corrective loop in charge to re-align the system while in closed loop. The top panel shows the output of the closed loop estimator while the bottom panel shows the actual position of the WFS translation stage.

\begin{figure}[b!] 
        \centering
        
        \newcommand{\PathFig}{Fig_misreg_CL_corr_GPAO/}

        \newcommand{\fontTxt}[1]{\textbf{\small #1}}        
        
        \newcommand{\LineRatio}{0.95}
        
        \newcommand{\subfigColor}{black}
        
        \newcommand{\figOne}{\PathFig Mis-registration_update}
        \newcommand{\figTwo}{\PathFig 2D_frequency_space}
        \newcommand{\figThree}{\PathFig temporal_frequency_space}
        
        \sbox1{\includegraphics{\figOne}}
        \sbox2{\includegraphics{\figTwo}}
        \sbox3{\includegraphics{\figThree}}
        
        \newcommand{\ColumnWidth}[1]
                {\dimexpr \LineRatio \linewidth * \AspectRatio{#1} / (\AspectRatio{1} + \AspectRatio{2} + \AspectRatio{3}) \relax
                }
        \newcommand{\ColumnGap}{\hspace {\dimexpr \linewidth /4 - \LineRatio\linewidth /4 }}

        \begin{tabular}{
                @{\ColumnGap}
                M{\ColumnWidth{1}}
                @{\ColumnGap}
                M{\ColumnWidth{2}}
                @{\ColumnGap}
                M{\ColumnWidth{3}}
                @{\ColumnGap}
                }
                
                \fontTxt{$\quad$ Lateral error update}
                &
                \fontTxt{2D spatial frequency space}
                &
                \fontTxt{Temporal frequency space}
                \\  
                \subfigimg[width=\linewidth,pos=ul,font=\fontfig{\subfigColor}]{$\quad\,\,$(a)}{0.0}{\figOne} &
                \subfigimg[width=\linewidth,pos=ul,font=\fontfig{\subfigColor}]{(b)}{0.75}{\figTwo} &
                \subfigimg[width=\linewidth,pos=ul,font=\fontfig{\subfigColor}]{(c)}{0.75}{\figThree}              
        \end{tabular}      
               
        \caption{\label{fig:CL_corr_GPAO} Monitoring and correcting the lateral error in the GPAO bench. \refpan{a}: convergence of the corrective loop. Top: lateral error estimated by the proposed method. Bottom: absolute positioning of the WFS stage. \refpans{b,c}: CDMS screenshots of the estimator sanity check outputs. \refpan{b}: Map of the 2D coefficients~$\corrt{2D}$. \refpan{c}: Curve of the temporal correlation~$\corrt{t}$ (red). The black curve is the theoretical correlation curve~$\Vcorr{0}$ for~$\Vdelta\rightarrow 0$.}
        
\end{figure}

First, it can be seen that the estimator found some lateral errors along the two axes, despite the WFS has been translated along a single axis. This is because there was an angle between the DM and the WFS, not lying in a Fried geometry\cite{Fried:77}. This angle is accounted for by the corrective loop.

Second, the amplitude estimated by the closed loop estimator was $\sim \percent{12}\SApitch$, well below the $\percent{60}\SApitch$ theoretically injected. As discussed in depth in \citet{Berdeu:24_misreg}, this is a known drawback of the closed loop estimator. Its sensitivity depends on the noise sources in the system, impacting the normalization in \refeq{eq:corr_empirical}. It is nonetheless always smaller than one, ensuring the stability of the lateral error corrective loop.

As shown by the bottom panel of \refsubfig{fig:CL_corr_GPAO}{a}, this corrective loop successfully converged. Bur there was a slight bias compared to the reference position. This reference was obtained with the phase plates stopped but inserted in the beam. They are known to introduce wobble in the system, adding an additional lateral error evolving with their rotation angle. The mentioned bias is considered to be within the margins of this wobble. It was thus considered that the estimator successfully converged without any strong bias induced by the frozen flow components of the turbulence.

In addition, the estimator has been integrated in SPARTA so that some sanity check outputs are published in its Configuration Data Management System (CDMS). Namely, the 2D map of the best fit of the coupling coefficients~$\theta$:
\begin{align}
    \label{eq:corr_2D}
    \corrt{2D}\Paren{\Vk} 
    {}\triangleq{} \argmin{\corr{}}{\sum_{f>0}\Paren{\corrt{cl}\Paren{\Vk,f} - \Vcorr{0}\Paren{f}\corr{}}^2}
    {}={} \frac{\sum_{f>0}\corrt{cl}\Paren{\Vk,f}\Vcorr{0}\Paren{f}}{\sum_{f>0} \Vcorr{0}^2\Paren{f}}
    \,,
\end{align}
shown in \refsubfig{fig:CL_corr_GPAO}{b}, and the best fit of the correlation curves:
\begin{equation}
    \label{eq:corr_t}
    \corrt{t}\Paren{f} 
    {}\triangleq{} \argmin{\corr{}}{\sum_{\Vk\in\kctrl}\Paren{\corrt{cl}\Paren{\Vk,f} - 2\pi\corr{\,}\Vk\scaprod\tilde{\Vdelta}}^2}
    {}={} \frac{\sum_{\Vk\in\kctrl}\corrt{cl}\Paren{\Vk,f}\Vk\scaprod\tilde{\Vdelta}}{\sum_{\Vk\in\kctrl} 2\pi\Paren{\Vk\scaprod\tilde{\Vdelta}}^2}
    \,,
\end{equation}
shown in \refsubfig{fig:CL_corr_GPAO}{c}.

Looking at \refsubfig{fig:CL_corr_GPAO}{b}, it appears that the map follows the expected `tip-tilt' pattern from \refeq{eq:theta} according to the 2D spatial frequency. In addition, the correlation is limited to the space controlled by the AO system, delimited by the green circle (dictated by the number of controlled modes, namely 800 here, see \citet{Berdeu:24_misreg} for more details).

Finally, the best fit of the temporal correlation curves of the command telemetry, shown in \refsubfig{fig:CL_corr_GPAO}{c}, nicely matches the theoretical curve, in black. At low temporal frequencies, their is a corruption of the correlation signal by the wind. As expected, this effect remains sparse, limited to a narrow frequency range. It did not impact the quality of the convergence of the lateral error corrective loop.

\subsubsection{First on-sky results}

In the previous section, the closed loop estimator was tested against phase plates emulating frozen flow turbulence, the worst case offender for the method. In reality, atmospheric turbulence is more complex, potentially presenting several layers and characteristic wind speeds or coherence times that may impact the estimator performances. Before the upcoming commissioning of GRAVITY+, it was thus interesting to test the method on data acquired on-sky with other instruments.

The CHARA collaboration kindly shared data acquired during technical time (2023-10-16) from the E2 unit. An additional `Garching remote access facility' (G-RAF) session (2024-05-28) on the GRAVITY/CIAO system, based on the soon to be decommissioned Multiple Application Curvature Adaptive Optics\cite{Arsenault:03_MACAO} (MACAO), was also granted by the European Southern Observatory on the third telescope unit (UT3) of the Very Large Telescope (VLT).

\begin{figure}[t!]
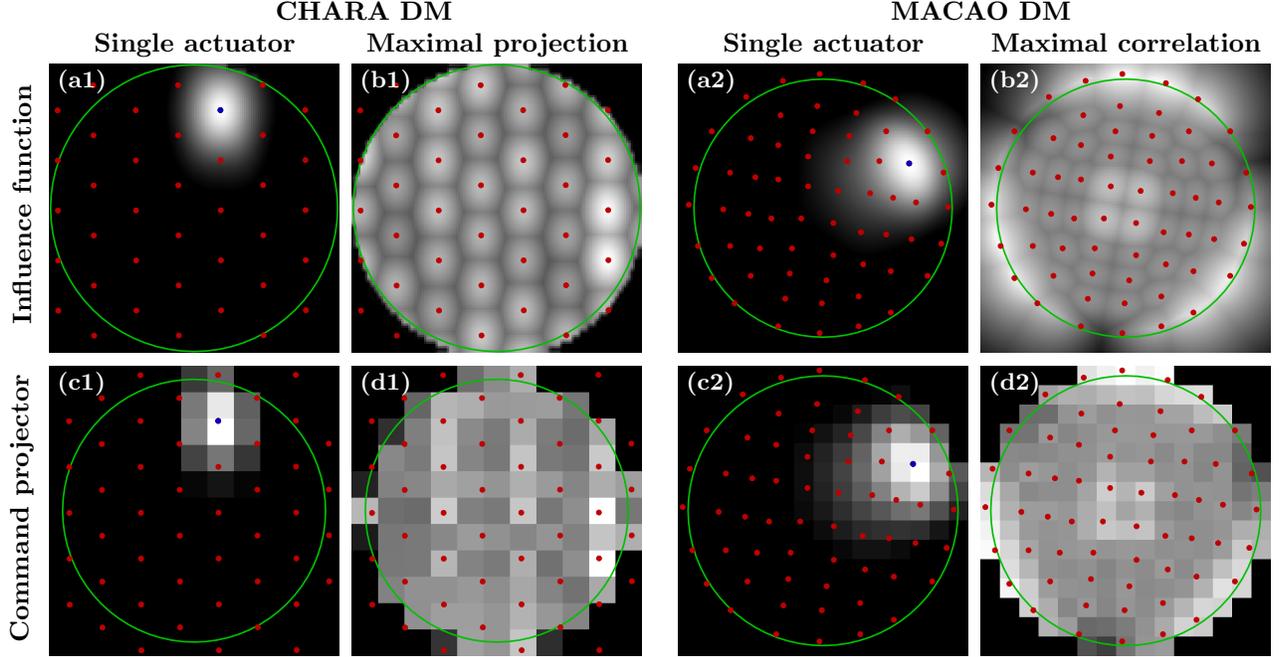
 
        \centering
        
        \newcommand{\PathFig}{Fig_misreg_CL_corr}
        
        \newcommand{\LineRatio}{0.925}
        
        \newcommand{\fontTxt}[1]{\textbf{#1}}
        
        \newcommand{\subfigColor}{white} 
        
        \newcommand{\widthTxt}{12pt}
        \newcommand{\widthFig}{\dimexpr (\linewidth - \widthTxt)}
        
        \newcommand{\ColumnWidth}
                {\dimexpr \LineRatio \widthFig / 4 \relax
                }
        \newcommand{\ColumnGap}{\hspace {\dimexpr \widthFig /8 - \LineRatio\widthFig /8 }}

        \begin{tabular}{
                @{\ColumnGap}
                M{\widthTxt}
                @{\ColumnGap}
                M{\ColumnWidth}
                @{\ColumnGap}
                M{\ColumnWidth}
                @{\ColumnGap}
                @{\ColumnGap}
                @{\ColumnGap}
                M{\ColumnWidth}
                @{\ColumnGap}
                M{\ColumnWidth}
                @{\ColumnGap}
                }
                
                &
                \multicolumn{2}{c}{\fontTxt{CHARA DM}}
                &
                \multicolumn{2}{c}{\fontTxt{MACAO DM}}
                \\[-0.1cm]
                &
                \fontTxt{Single actuator}
                &
                \fontTxt{Maximal projection}
                &
                \fontTxt{Single actuator}
                &
                \fontTxt{Maximal correlation}
                \\
                \rotatebox[origin=l]{90}{\fontTxt{Influence function}} &
                \subfigimg[width=\linewidth,pos=ul,font=\fontfig{\subfigColor}]{$\,$(a1)}{0.0}{\PathFig _CHARA/DM_act_35} &
                \subfigimg[width=\linewidth,pos=ul,font=\fontfig{\subfigColor}]{$\,$(b1)}{0.0}{\PathFig _CHARA/DM_proj} &
                \subfigimg[width=\linewidth,pos=ul,font=\fontfig{\subfigColor}]{$\,$(a2)}{0.0}{\PathFig _CIAO/DM_act_35} &
                \subfigimg[width=\linewidth,pos=ul,font=\fontfig{\subfigColor}]{$\,$(b2)}{0.0}{\PathFig _CIAO/DM_proj}
                \\
                \rotatebox[origin=l]{90}{\fontTxt{Command projector}} &
                \subfigimg[width=\linewidth,pos=ul,font=\fontfig{\subfigColor}]{$\,$(c1)}{0.0}{\PathFig _CHARA/mod_act_35} &
                \subfigimg[width=\linewidth,pos=ul,font=\fontfig{\subfigColor}]{$\,$(d1)}{0.0}{\PathFig _CHARA/mod_proj} &
                \subfigimg[width=\linewidth,pos=ul,font=\fontfig{\subfigColor}]{$\,$(c2)}{0.0}{\PathFig _CIAO/mod_act_35} &
                \subfigimg[width=\linewidth,pos=ul,font=\fontfig{\subfigColor}]{$\,$(d2)}{0.0}{\PathFig _CIAO/mod_proj}
        \end{tabular}        
               
        \caption{\label{fig:CL_coor_DM_model} Non-Cartesian grids of the CHARA (\refpans{1}) and GRAVITY/CIAO DM (\refpans{2}). \refpans{a,b}: High resolution influence function and their maximal projection. \refpans{c,d}: Low resolution model for the command 2D projection. Green circle: pupil edge. Colored dots: actuator position.
        }
\end{figure}

The DMs of both systems own 60 actuators. Nonetheless, they do not lie on a Cartesian grid, as shown on \refsubfigs{fig:CL_coor_DM_model}{a,b}. As a consequence, the commands from the AO telemetry cannot be directly linked with a 2D representation. To overcome this issue, a low resolution projector, with $\sim 1.5$ pixel per actuator, was used to reshape the DM commands in 2D patterns, as presented in \refsubfigs{fig:CL_coor_DM_model}{c,d}.

The main parameters impacting the correlation curves of the closed loop estimator are listed in \reftab{tab:AO_parameters}. For each instrument, a batch of 2500 telemetry frames was used, corresponding to only \SI{5}{\second}.

\begin{table}[h!] 
    \caption{\label{tab:AO_parameters} Parameters of the CHARA and CIAO AO loops.}
    \centering
    \begin{tabular}{ccc}
    \hline
    \hline
     & CHARA & CIAO
    \\
    \hline
    Loop frequency $1/\tWFS$ (\SI{}{Hz}) & 441 & 500
    \\
    Loop gain $\gint$ & 0.19 & 0.4
    \\
    Number of controlled mode & 41 & 45
    \\
    \hline
    \end{tabular}
\end{table}

\begin{figure}[t!]
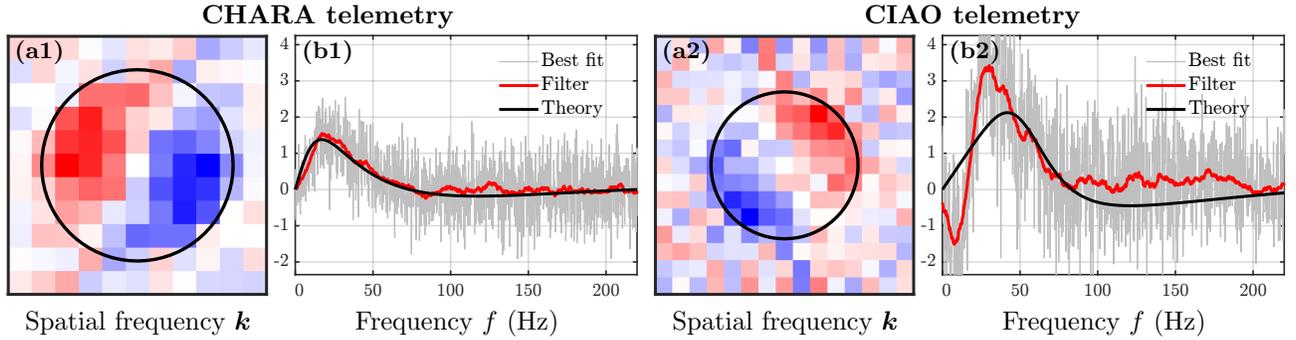
 
        \centering
        
        \newcommand{\PathFig}{Fig_misreg_CL_corr}
        
        \newcommand{\LineRatio}{0.975}
        
        \newcommand{\fontTxt}[1]{#1}
        
        \newcommand{\subfigColor}{black} 
                
		\newcommand{\figOne}{\PathFig _CHARA/corr_map}
        \newcommand{\figTwo}{\PathFig _CHARA/corr_curv}
        \newcommand{\figThree}{\PathFig _CIAO/corr_map}
        \newcommand{\figFour}{\PathFig _CIAO/corr_curv}

        \sbox1{\includegraphics{\figOne}}
        \sbox2{\includegraphics{\figTwo}}
        \sbox3{\includegraphics{\figThree}}
        \sbox4{\includegraphics{\figFour}}

        \newcommand{\ColumnWidth}[1]
                {\dimexpr \LineRatio \linewidth * \AspectRatio{#1} / (\AspectRatio{1} + \AspectRatio{2} + \AspectRatio{3}+ \AspectRatio{4}) \relax
                }
        \newcommand{\ColumnGap}{\hspace {\dimexpr \linewidth /7 - \LineRatio\linewidth /7 }}

        \begin{tabular}{
                @{\ColumnGap}
                M{\ColumnWidth{1}}
                @{\ColumnGap}
                M{\ColumnWidth{2}}
                @{\ColumnGap}
                @{\ColumnGap}
                @{\ColumnGap}
                M{\ColumnWidth{3}}
                @{\ColumnGap}
                M{\ColumnWidth{4}}
                @{\ColumnGap}
                }
                
                \multicolumn{2}{c}{\textbf{CHARA telemetry}}
                &
                \multicolumn{2}{c}{\textbf{CIAO telemetry}}
                \\
                \subfigimg[width=\linewidth,pos=ul,font=\fontfig{\subfigColor}]{$\,$(a1)}{0.0}{\figOne} &
                \subfigimg[width=\linewidth,pos=ul,font=\fontfig{\subfigColor}]{$\quad\!$(b1)}{0.0}{\figTwo} &
                \subfigimg[width=\linewidth,pos=ul,font=\fontfig{\subfigColor}]{$\,$(a2)}{0.0}{\figThree} &
                \subfigimg[width=\linewidth,pos=ul,font=\fontfig{\subfigColor}]{$\quad\!$(b2)}{0.0}{\figFour}
                \\[-0.1cm]
                \fontTxt{Spatial frequency $\Vk$}
                &
                \fontTxt{Frequency $f$ (\SI{}{Hz})}
                &
                \fontTxt{Spatial frequency $\Vk$}
                &
                \fontTxt{Frequency $f$ (\SI{}{Hz})}
                \\[0.1cm]
        \end{tabular}        
               
        \caption{\label{fig:CL_corr_sky} Sanity check outputs of the lateral error estimator with on-sky closed loop telemetry from CHARA (\refpans{1}) and GRAVITY/CIAO (\refpans{2}). \refpans{a}: Map of the 2D coefficients~$\corrt{2D}$. The black circles encompass the spatial frequencies of the controlled space. \refpans{b}: Curve of the temporal correlation~$\corrt{t}$ (gray). The black curve is the theoretical correlation curve~$\Vcorr{0}$ for~$\Vdelta\rightarrow 0$. The red curve is a mean filter of~$\Vcorrt{t}$ with a sliding window of $\pm25$ data points. 
        }
\end{figure}

The sanity check outputs of the closed loop estimator are given in \reffig{fig:CL_corr_sky}. As for the results in the GPAO bench, the `tip-tilt' pattern is nicely visible in the maps of the coupling coefficients~$\corrt{2D}$ and is well encompassed in the predicted controlled area, see \refsubfigs{fig:CL_corr_sky}{a1,a2}.  Looking at the temporal correlation curves, see \refsubfigs{fig:CL_corr_sky}{b1,b2}, it first appears that the theoretical curves, in black, are different. This comes from the different gains of the AO loop used in each system. Despite the small number of actuators and controlled modes\footnote{60 actuators and $\leq$ 45 controlled modes. To be compared with the 1432 active actuators of GPAO, controlling 500 to 800 modes...} and the shortness of the gathered telemetry, the correlation signal is clear in the gray curves. Once filtered with a sliding windows, in red, the curve perfectly matches the prediction in the CHARA dataset. In the CIAO telemetry, the low frequencies are polluted by some wind signal.

\begin{figure}[t!]
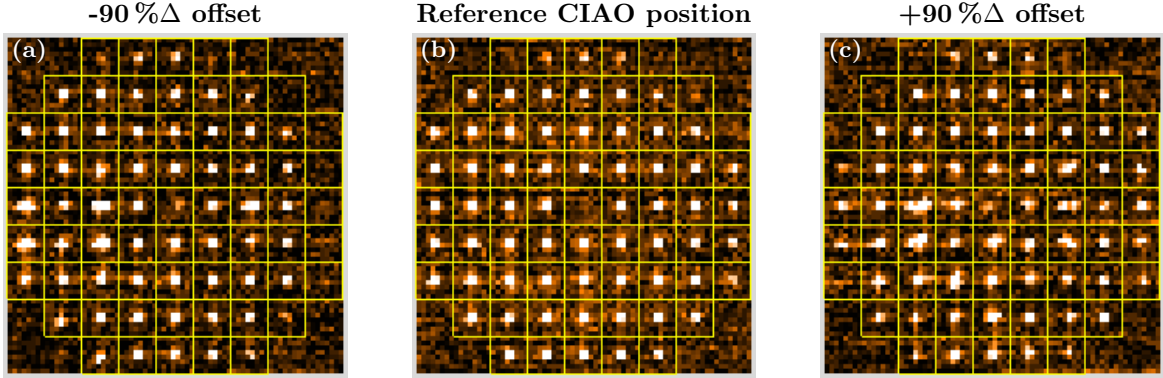
 
        \centering
        
        \newcommand{\PathFig}{Fig_misreg_CIAO_misreg/Misreg_ND_x}

        \newcommand{\fontTxt}[1]{\textbf{#1}}        
        
        \newcommand{\LineRatio}{0.8}
        
        \newcommand{\subfigColor}{white}
        
        \newcommand{\ColumnWidth}
                {\dimexpr \LineRatio \linewidth / 3 \relax
                }
        \newcommand{\ColumnGap}{\hspace {\dimexpr \linewidth /4 - \LineRatio\linewidth /4 }}

        \begin{tabular}{
                @{\ColumnGap}
                M{\ColumnWidth}
                @{\ColumnGap}
                M{\ColumnWidth}
                @{\ColumnGap}
                M{\ColumnWidth}
                @{\ColumnGap}
                }
                
                \fontTxt{\percent{-90}$\SApitch$ offset}
                &
                \fontTxt{Reference CIAO position}
                &
                \fontTxt{\percent{+90}$\SApitch$ offset}
                \\  
                \subfigimg[width=\linewidth,pos=ul,font=\fontfig{white}]{$\,$(a)}{0.0}{\PathFig -150} &
                \subfigimg[width=\linewidth,pos=ul,font=\fontfig{white}]{$\,$(b)}{0.0}{\PathFig +000} &
                \subfigimg[width=\linewidth,pos=ul,font=\fontfig{white}]{$\,$(c)}{0.0}{\PathFig +150}              
        \end{tabular}

%
%
%
%
%
%
               
        \caption{\label{fig:CL_corr_CIAO_bias} Application of different lateral offsets along the $x$-axis on the CIAO WFS stage. Loop closed on the VLT UT3 internal beacon inserting a neutral density filter in the beam.}
\end{figure}

To further investigate the impact of real turbulence wind on the closed loop estimator, known lateral shifts were introduced in the CIAO system, as shown in \reffig{fig:CL_corr_CIAO_bias}. The CIAO WFS was translated by steps of \SI{25}{\micro\meter}, up to \SI{\pm150}{\micro\meter}, corresponding to an misalignment of $\percent{\pm90}\SApitch$.

The estimator was first tested on the beacon of the VLT UT3. To maximize the mis-registration signal, induced by the noise propagation through the AO loop, a neutral density filter was inserted in the beam. For each shift, \SI{20}{\second} of telemetry was recorded (10000 points). To get some statistics and assess the estimator covariance, 31 batches of 2500 consecutive frames every 250 telemetry frames were analyzed.

\begin{figure}[t!]
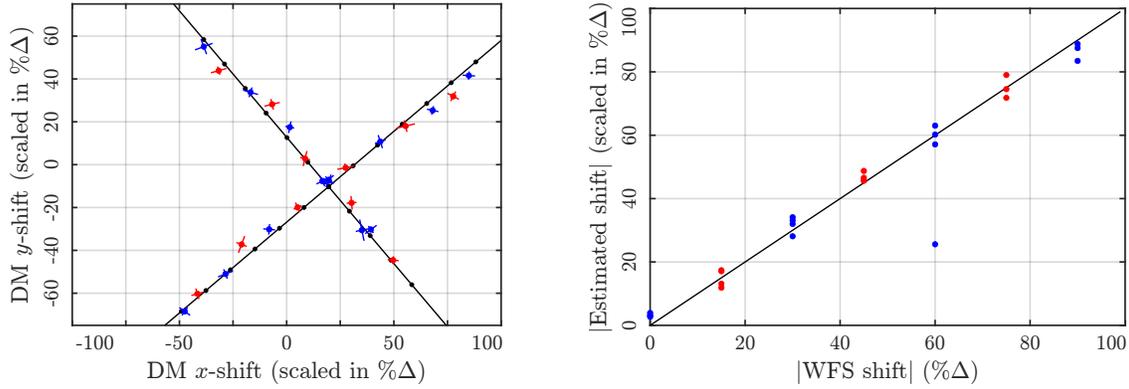
 
        \centering
        
        \newcommand{\PathFig}{Fig_misreg_CIAO_gain_CIAO_beacon_ND/}

        \newcommand{\LineRatio}{0.8}
        
        \newcommand{\subfigColor}{black}
        
        \newcommand{\figOne}{\PathFig Gain_2D}
        \newcommand{\figTwo}{\PathFig Gain}

        \sbox1{\includegraphics{\figOne}}
        \sbox2{\includegraphics{\figTwo}}

        \newcommand{\ColumnWidth}[1]
                {\dimexpr \LineRatio \linewidth * \AspectRatio{#1} / (\AspectRatio{1} + \AspectRatio{2}) \relax
                }
        \newcommand{\ColumnGap}{\hspace {\dimexpr \linewidth /3 - \LineRatio\linewidth /3 }}
               
        \begin{tabular}{
                @{\ColumnGap}
                M{\ColumnWidth{1}}
                @{\ColumnGap}
                M{\ColumnWidth{2}}
                @{\ColumnGap}
                }
                \subfigimg[width=\linewidth,pos=ul,font=\fontfig{white}]{}{0.0}{\PathFig Gain_2D} &
                \subfigimg[width=\linewidth,pos=ul,font=\fontfig{white}]{}{0.0}{\PathFig Gain}              
        \end{tabular}   
               
        \caption{\label{fig:CL_corr_CIAO_bias_beacon} Comparison on the VLT/UT3 beacon of the closed loop estimator (blue and red alternates every \percent{15}$\SApitch$) with known lateral mis-registrations applied on the GRAVITY/CIAO WFS (back curves and dots). \refpan{a}: 2D estimated lateral errors scaled to match the applied shifts in order to compensate for the estimator sensitivity. \refpan{b}: Norm of the estimated lateral error in function of the applied shift (corrected for the bias of \refpan{a}).} 
        
\end{figure}

The results are shown in \refsubfig{fig:CL_corr_CIAO_bias_beacon}{a}. The bars indicate the $1\sigma$ error axes derived from the 2D covariances. It first appears that the estimated mis-registrations are rotated compared to the theoretical lateral shifts applied along the WFS $x$ and $y$-axes. This clocking is induced by the K-mirror position in-between the MACAO DM and the CIAO WFS. As mentioned above, the sensitivity of the closed loop estimator is not equal to unity and the comparison of its outputs with the injected shifts is not straightforward. To do so, the best linear transform linking the estimated mis-registrations~$\tilde{\Vdelta}$  with their corresponding theoretical lateral shifts~$\Vdelta^\Tag{th}$ was computed as follows:
\begin{equation}
    \Paren{\tilde{\Vdelta_{0}}, \tilde{\rho}, \tilde{\alpha}} 
    {}={} \argmin{\Vdelta_{0}, \rho, \alpha}{\sum_{\Vdelta^\Tag{th}}\Paren{\tilde{\Vdelta}\paren{\Vdelta^\Tag{th}} - \rho\times\Paren{\Vdelta_{0} + \V{R}\Paren{\alpha}\times\Vdelta^\Tag{th}}}^2}
    \,,
\end{equation}
where $\V{R}\Paren{\alpha}$ is the rotation matrix of angle~$\alpha$. In \refsubfig{fig:CL_corr_CIAO_bias_beacon}{a}, the estimated lateral errors were thus scaled by $1/\rho$, to be put in regards to the injected shifts rotated in the DM space by $\alpha$ (black dots).

It then appears that there is a bias of $\Vdelta_0 = \percent{\Paren{19.5, -10.3}}\SApitch$. The origin of this bias is not clear. Indeed, in absence of strong frozen flow, the estimator should not be biased, an estimated shift of $\tilde{\Vdelta}=\V{0}$ then corresponding to a system without any lateral mis-registration. This bias was attributed to a mis-alignment of the CIAO WFS reference, potentially produced by the strong vignetting of the pupil, see \refsubfig{fig:CL_corr_CIAO_bias}{b}.

\refsubfigfull{fig:CL_corr_CIAO_bias_beacon}{b} shows the norm of the estimated shift in function of the injected shift, after correcting for the bias $\tilde{\Vdelta}=\V{0}$. The estimator appears linear up to the tested range of $\percent{90}\SApitch$, except for a point at $\percent{60}\SApitch$ (visible in \refsubfig{fig:CL_corr_CIAO_bias_beacon}{a}, at $\sim\percent{\Paren{30,-30}}\SApitch$). This point is actually at the limit of the loop stability of the system, degrading the empirical variance estimate and thus the fit of the closed loop estimator. Divergence occurring for larger mis-registrations in this direction whereas the loop stays stable in other directions can be linked with the pupil vignetting but also can support the fact that there is indeed a bias in the lateral reference of the CIAO WFS.

The same tests were reproduced on-sky, targeting a star of magnitude 7, with an approximated seeing of 0.6''. For each injected lateral shift, \SI{10}{\second} of telemetry was recorded (5000 points). These telemetries were analyzed through 11 batches of 2500 consecutive frames every 250 telemetry frames. The results are shown in \reffig{fig:CL_corr_CIAO_bias_beacon}. Contrary to the results on the VLT/UT3 beacon, the cross pattern is slightly distorted for high mis-registration value. As expected, the error bars indicate that the error is mainly along the direction of the induced shift. Nonetheless, the pattern of the estimated shifts clearly presents an origin where a corrective loop based on the closed estimator would converge. With an offset of $\Vdelta_0 = \percent{\Paren{-8.0, -5.3}}\SApitch$, its amplitude is consistent with the results obtained on the beacon. This result once again suggests that no evident bias is induced by the turbulence.

\begin{figure}[t!]
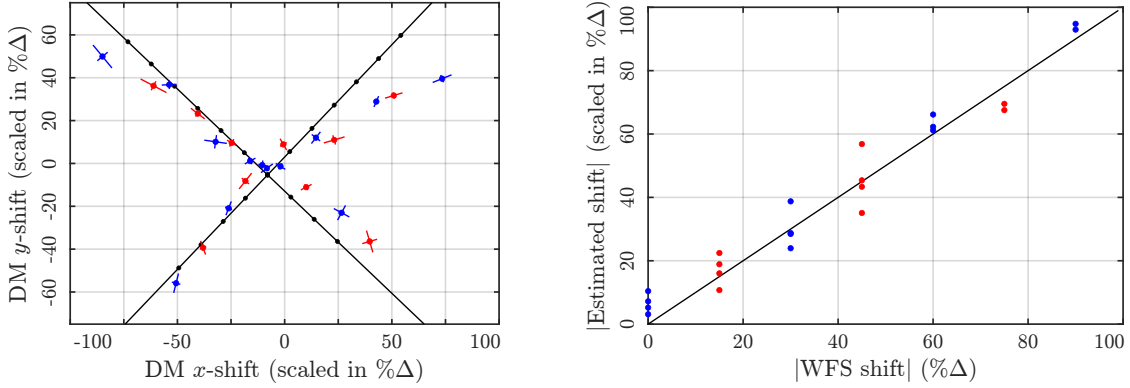
 
        \centering
        
        \newcommand{\PathFig}{Fig_misreg_CIAO_gain_CIAO_on_sky/}

        \newcommand{\LineRatio}{0.8}
        
        \newcommand{\subfigColor}{black}
        
        \newcommand{\figOne}{\PathFig Gain_2D}
        \newcommand{\figTwo}{\PathFig Gain}

        \sbox1{\includegraphics{\figOne}}
        \sbox2{\includegraphics{\figTwo}}

        \newcommand{\ColumnWidth}[1]
                {\dimexpr \LineRatio \linewidth * \AspectRatio{#1} / (\AspectRatio{1} + \AspectRatio{2}) \relax
                }
        \newcommand{\ColumnGap}{\hspace {\dimexpr \linewidth /3 - \LineRatio\linewidth /3 }}
               
        \begin{tabular}{
                @{\ColumnGap}
                M{\ColumnWidth{1}}
                @{\ColumnGap}
                M{\ColumnWidth{2}}
                @{\ColumnGap}
                }
                \subfigimg[width=\linewidth,pos=ul,font=\fontfig{white}]{}{0.0}{\PathFig Gain_2D} &
                \subfigimg[width=\linewidth,pos=ul,font=\fontfig{white}]{}{0.0}{\PathFig Gain}              
        \end{tabular}   
               
        \caption{\label{fig:CL_corr_CIAO_bias_on_sky} Same as Fig.~\ref{fig:CL_corr_CIAO_bias_beacon} but on a 7 magnitude star with good seeing condition (0.6'').}
        
\end{figure}

Finally, let's note once again that \reffigs{fig:CL_corr_CIAO_bias_beacon}{fig:CL_corr_CIAO_bias_on_sky} do not emphasize the variability of the closed loop estimator sensitivity. As already discussed above, it is impacted by the normalization by the variances in \refeq{eq:corr_empirical} that may decrease the estimated empirical covariance. For these two examples, the sensitivity $\rho$ is \percent{10} smaller in presence of turbulence with a bright star than with the beacon after the neutral density filter.


\section{Conclusions and perspectives}

In this work, we introduced two novel methods to estimate the lateral misalignment of an AO system that relax the need of complex PSIM model fitting. After a brief reminder of their underlying theory, they were experimentally validated in the GPAO development bench and with on-sky data obtained from current AO facilities. These methods will be at the core of the GPAO system auto-alignment strategy while in operation to keep the AO performances at their optimal level throughout the observations.

\subsection{Perturbative 2D Modal Estimator}

Contrary to other solutions based on modal perturbations\cite{Heritier:21_SPRINT}, this method does not imply the fine tuning of the PSIM model parameters but uses the 2D representation of the IMs to directly estimate the lateral errors. This method has a large capture range, is unbiased (absolute) and robust to large misalignment errors and low S/Ns. In addition, the method is fast: (i) the acquisition of the partial modal IM is quick to obtain, and (ii) the convolutions in \refeq{eq:2D_corr_map} can be computed using fast 2D discrete Fourier transforms and the fit of the lateral error is obtained in a single pass via \refeq{eq:2D_corr_estim}, not relying on an iterative fit.

Furthermore, the need of a PSIM model is reduced to the minimal need: a single computation to get the reference IM from which the command matrix of the AO loop is derived. For complex systems where a PSIM model is not available, this need could be removed by using a measured IM obtained during calibration on an aligned system.

This method is thus particularly suitable for open-loop calibration where the turbulence strongly corrupts the slopes of the measured modal IM, as far as the measurement strategy freezes, in average, the turbulence\cite{Kolb:16_SPIE_Review_AO_calibration}. In addition, \citet{Heritier:21_SPRINT} has shown that these kinds of approaches are not biaised by the turbulence wind.

\subsection{Non-Perturbative Closed Loop Estimator}

One of the main advantages of the closed loop estimator is that it is purely based on the geometry of the chosen observable (the WFS measurements~$\Vm_{i}$ or the DM commands~$\Vc_{i}$). This relaxes the need of PSIM models, potentially complex and hard to fit, from which mis-registration parameters must be retrieved.

Secondly, the estimator depends on a limited number of parameters driving the AO loop. They are generally well constrained by design or proper calibration.

Furthermore, the estimator, expressed in the Fourier domain, is sparse. This makes it extremely fast to compute, a critical aspect when considering the ever increasing number of actuators and system complexity of the future ELTs.

Even if it based on noise propagation, assumptions on the noise model remain minimal. It only supposes that the noise is spatially uncorrelated.

In addition, a common issue with mis-registration estimators is their sensitivity to frozen flow turbulence\cite{Heritier:19_frozen_wind, Heritier:19_PHD}. As discussed in \citet{Berdeu:24_misreg}, such a situation breaks the assumption that~$\Vp_{1}$ and~$\Vp_{2}$ are independent, used to get \refeq{eq:corr_0}. Nonetheless, other sources of noise are always injected in the loop, producing a correlation signal in the commands. As discussed in \ref{sec:meth_CL_estim}, the frozen flow signal then becomes sparse in the temporal frequency space. Thus, compared to other methods based on AO loop telemetry\cite{Bechet:12_SPIE, Kolb:12_SPIE_AOF}, the wind has a limited impact on the proposed estimator of \refeq{eq:CL_estim}, as shown by the results obtained on-bench and on-sky.

Finally, one should keep in mind that this estimator is only relative: it is sensitive to a mis-registration between the real AO system and its numerical model used to close it. It makes only sense in the context of lateral error monitoring and correction and cannot be used to assess absolute lateral mis-registrations.

\subsection{What's Next?}

In the short term, the upcoming commissioning of GRAVITY+ and its brand new AO system will be the occasion to further test and characterize these new lateral estimators in real conditions, with a large variety of seeings and turbulence strengths. This may raise the need to further work on the debiasing of the closed loop estimator, as we saw that the wind produces a well identified signal in the telemetry correlation. This commissioning will also the the occasion to confirm these algorithms as current baseline for the upcoming European ELT\cite{Bonnet:23_AO4ELT_ELT}.

Work is currently on-going to extend these methods to higher order mis-registrations, namely the clocking, the magnification and anamorphoses. This nonetheless implies more complex block diagrams in the spatial frequency coupling than the one introduced in \refsubfig{fig:CL_corr_curves}{a}.

In a longer term, these algorithms need to be confronted to other kinds of WFSs, such as the pyramid. In this context, their adaptation is relatively straightforward, as a pyramid WFS directly produces images of spatial features in the instrument pupil. The close loop estimator performs well in simulation but is yet to be confirmed in the lab or on-sky with such WFSs.

\acknowledgments 
 
This project has received funding from the European Un{\-}ion's Horizon 2020 research and innovation programme under grant agreement No 101004719.

This work made use of data acquired during CHARA and VLTI technical time.

\bibliography{2024_SPIE_Berdeu-et-al} 
\bibliographystyle{spiebib} 

\end{document}